\documentclass[aps,prl,twocolumn,showpacs,preprintnumbers,amsmath,amssymb]{revtex4-2}
\usepackage{bm}
\usepackage{amsmath,amssymb,amsthm,times,graphics,graphicx,bm,xcolor}
\usepackage[normalem]{ulem}

\usepackage{xcolor}
\usepackage{bbold}

\newcommand{\ket}[1]{\left | #1 \right\rangle}
\newcommand{\bra}[1]{\left \langle #1 \right |}

\begin{document}


\title{Emergence of Constructor-based Irreversibility in Quantum Systems:\\ Theory and Experiment}

\author{Chiara Marletto and Vlatko Vedral}
\affiliation{Clarendon Laboratory, University of Oxford, Parks Road, Oxford OX1 3PU, United Kingdom and\\ Fondazione ISI, Via Chisola 5, Torino, Italy and \\
Centre for Quantum Technologies, National University of Singapore, 3 Science Drive 2, Singapore 117543 and\\
Department of Physics, National University of Singapore, 2 Science Drive 3, Singapore 117542}
\author{Laura Knoll}
\affiliation{Istituto Nazionale di Ricerca Metrologica, Strada delle Cacce 91, 10135, Torino, Italy}
\author{Fabrizio Piacentini}
\affiliation{Istituto Nazionale di Ricerca Metrologica, Strada delle Cacce 91, 10135, Torino, Italy}
\author{Ettore Bernardi}
\affiliation{Istituto Nazionale di Ricerca Metrologica, Strada delle Cacce 91, 10135, Torino, Italy}
\author{Enrico Rebufello}
\affiliation{Istituto Nazionale di Ricerca Metrologica, Strada delle Cacce 91, 10135, Torino, Italy}
\author{Alessio Avella}
\affiliation{Istituto Nazionale di Ricerca Metrologica, Strada delle Cacce 91, 10135, Torino, Italy}
\author{Marco Gramegna}
\affiliation{Istituto Nazionale di Ricerca Metrologica, Strada delle Cacce 91, 10135, Torino, Italy}
\author{Ivo Pietro Degiovanni}
\author{Marco Genovese}
\affiliation{Istituto Nazionale di Ricerca Metrologica, Strada delle Cacce 91, 10135, Torino, Italy and, INFN, sezione di Torino, via P. Giuria 1, 10125 Torino, Italy}

\date{\today}



\begin{abstract}

The issue of irreversibility in a universe with time-reversal-symmetric laws is a central problem in physics. 
In this letter, we discuss for the first time how irreversibility can emerge within the recently proposed constructor theory framework. Here irreversibility is expressed as the requirement that a task is possible, while its inverse is not.
In particular, we demonstrate that this irreversibility is compatible with quantum theory's time reversal symmetric laws, by exploiting a specific model, based on the universal quantum homogeniser, realised experimentally with high-quality single-photon qubits.

\end{abstract}

\maketitle

\textit{Introduction.} The emerging of irreversibility from time-symmetric physical laws is a central problem in contemporary physics.
Indeed, there are several approaches to irreversibility in physics: statistical mechanics methods \cite{BAY, BUC, WAL1}; information-theoretic descriptions of logically irreversible tasks, such as reset \cite{LAN, BEN, EAR}; and classical and quantum thermodynamics second laws \cite{BUC, THD1, THD2, THD3}. In all such cases, a tension arises between the laws that describe the irreversible phenomena, and the time-reversal symmetric laws that describe the microscopic constituents.
In this paper we focus on a particular type of irreversibility, which we express as the requirement that a transformation is possible (i.e., there is no limitation to how well it can be realised by a system operating in a cycle), while its inverse is not.
Joule's experiment \cite{BUC} in classical thermodynamics is an example where this irreversibility manifests itself: while it is possible to heat up a volume of water by mechanical means only, for instance with a stirrer, it is impossible to cool it down by the same means.
First introduced via a thermodynamic cycle (e.g. Carnot's), the idea of a cycle performing a transformation was generalised by von Neumann to a {\sl constructor} - a system that can perform a given task on another system and crucially retains the ability to perform the task again. Therefore, we call the irreversibility generalising that of Joule's experiment ``{\sl constructor-based} irreversibility". To analyse it formally, we use {\sl constructor theory}, a recently proposed generalisation of the quantum theory of computation to cover general tasks, \cite{DEU, DEMA}.
This irreversibility presents many advantages with respect to traditional approaches in (quantum) thermodynamics.
With respect to the usual information-theoretic description, it does not suffer from the circularity between the definition of information and distinguishability \cite{DEMA}.
Compared with traditional statistical mechanics methods, constructor-based irreversibility does not require the existence of dynamical trajectories in the phase space, difficult to define in a quantum framework.
Finally, concerning the second law of quantum thermodynamics, this irreversibility overcomes the limitations of the Two-Point Measurement scheme definition of work \cite{Perarnau-Llobet_2014}, i.e. the destruction of quantum coherences between system and thermal reservoir.
Starting from the general description of constructor-based irreversibility, we then specialise to the case of a qubit-based toy model where we demonstrate that, surprisingly, this type of irreversibility is compatible with quantum theory time-reversal symmetric laws, demonstrating it in an experiment exploiting high-quality photonic qubits.
Let us consider a universe made of qubits, available in unbounded numbers and in any state, where all unitary transformations and their transposes are allowed.
A task $T$ is the specification of a physical transformation on qubits, e.g. from a quantum state $\rho_x$ to another one $\rho_y$:

\begin{equation}
T=\{\rho_x \rightarrow \rho_y\} \label{T}
\end{equation}
whose transpose $T^{\sim}$ is simply defined as:

\begin{equation}
T^{\sim}=\{\rho_y \rightarrow \rho_x\} \;.
\end{equation}

We will refer to the {\sl substrate} qubit on which the task $T$ is defined as $Q$, and to the {\sl rest} of the qubits as $R$.
A  {\sl constructor}  for the task ${T}$ on the substrate $Q$ is some subsystem of the rest enabling ${T}$, {\sl without undergoing any net change in its ability to do it again}.
A task is {\sl possible} if the laws of physics do not put any constraint on the accuracy to which it can be performed by a constructor. It is {\sl impossible} otherwise.

The main point of our analysis is that constructor-based irreversibility can be defined as the fact that, while the task $T$ is possible, its transpose $T^{\sim}$ is not. The second law of thermodynamics can also be expressed via a statement of this kind \cite{UFF}: this is a long-standing tradition in thermodynamics, initiated by Planck \cite{UFF} and continuing with axiomatic thermodynamics \cite{CAR,LIE}.
We express now the conditions for a {constructor} for the task $T$ to be allowed under our unitary quantum model.
A unitary transformation acting on a substrate $Q$ and the rest $R$ will be denoted by $\mathcal{U}_{Q,R}$.
For a fixed task $T$ on $Q$ and a positive $\epsilon$, define the set of quantum states of $R$ that can perform $T$ to accuracy $\epsilon$:
\begin{equation}\label{VeT}
V_{\epsilon}[T]\doteq\{\rho_R\;:\; \mathcal{U}_{Q,R}(\rho_x\otimes\rho_R)\mathcal{U}_{Q,R}^{\dagger}= \rho \;,\;{\rm Tr}_R[\rho]\in\epsilon(\rho_y)\}
\end{equation}
where $\epsilon(\rho_y)$ is the $\epsilon$-ball centered around $T$'s desired output state, $\rho_y$: $\epsilon(\rho_y)\doteq \{\sigma: F(\rho_y, \sigma) \geq 1-\epsilon$\}, and $F(\rho_\alpha, \rho_\beta)=({\rm Tr}(\sqrt{\sqrt{\rho_\alpha}\rho_\beta\sqrt{\rho_\alpha}}))^2$ is the quantum fidelity \cite{fidn}.
We shall denote with ${\cal E}[T]$ a set of $N$ qubits prepared in a state belonging to $V_{\epsilon}$, i.e. a ``machine'' capable of performing the task $T$ with an error $\epsilon$.

Let us now introduce a measure of how much a given ${\cal E}[T]$ can perform $T$ to accuracy $\epsilon$ after $n$ repeated usages on $n$ fresh substrate qubits $Q_1, ..., Q_n$, each in the input state $\rho_x^{(n)}=\rho_x\;\forall n$.
%
%
Define, for a given initial state of the rest $\rho_R$, a recursive expression for the rest density operator after the $n$-th usage of the machine:
%
%
\begin{eqnarray}\label{rhoRn}
  \nonumber \rho_R^{(1)}&=&{\rm Tr}_{Q_1}\left[\mathcal{U}_{Q_{1}, R}\left(\rho_x^{(1)}\otimes\rho_R\right)\mathcal{U}_{Q_1, R}^{\dagger}\right] \\
  &\vdots& \\
  \nonumber \rho_R^{(n)}&=&{\rm Tr}_{Q_n}\left[\mathcal{U}_{Q_{n}, R}\left(\rho_x^{(n)}\otimes\rho_R^{(n-1)}\right)\mathcal{U}_{Q_n, R}^{\dagger}\right]
\end{eqnarray}
where $\mathcal{U}_{Q_n,R}=U_{Q_n,N}\cdots U_{Q_n,1}$ denotes a sequence of unitary interactions between the $n$-th substrate qubit undergoing $T$ and the $N$ rest qubits emerging from the $(n-1)$-th task execution.\\
We can then define the worst-case-scenario steadiness of the machine ${\cal E}[T]$ after performing the task $n$ times:

\begin{equation}\label{delta}
 S_{{\cal E}[T]}(n)\doteq\displaystyle{{\rm Inf}_{\rho_R\in V_{\epsilon}[T]}}\{F(\rho_R,\rho_R^{(n)})\}\;.
\end{equation}

Most machines lose the ability to perform the task with repeated use, so we expect $S_{{\cal E}[T]}(n)$ to decrease with $n$ for a fixed $\epsilon$.
A figure of merit for the resiliency of ${\cal E}[T]$ is the {\sl relative deterioration} of ${\cal E}[T]$ after having been used $n$ times, defined as:
\begin{equation}\label{reldet}
 \delta_{\mathcal{E}[T]}(n)\doteq \frac{\epsilon}{S_{\mathcal{E}(T)}(n)}\;.
\end{equation}
%
%
There are two conditions for a constructor realising $T$ to be allowed under a given unitary law $\mathcal{U}_{Q_n,R}$.\\

{\sl Condition (i).}
{For any positive arbitrarily small $\epsilon$, the set $V_{\epsilon}[T]$ of Eq. (\ref{VeT}) is non-empty (i.e. the rest can perform the task $T$ to arbitrarily high accuracy, once.)} \\

{\sl Condition (ii).}
The relative deterioration $\delta_{\mathcal{E}[T]}(n)$ of the rest $R$ in its ability to perform the task $T$ with error $\epsilon$ goes to zero, as the error $\epsilon$ goes to zero and the number $n$ of repeated usages goes to infinity:
\begin{equation}\label{limR}
 \lim_{\epsilon \rightarrow 0}\lim_{n\rightarrow \infty}\delta_{\mathcal{E}[T]}(n)=0\;.
\end{equation}

Note that the order of the limits is relevant for the correct physical interpretation: for a fixed accuracy $\epsilon$, we let the number of usages go to infinity.
%

If these two conditions are both satisfied, then a sequence of machines ${\cal E}[T]$ converges to a limiting machine that perfectly retains the ability to cause the transformation with asymptotically small error, even when used $n$ times, for arbitrarily large $n$.
The limiting point of the sequence of machines $\{{\cal E}[T]\}$ is a constructor for the task $T$ (please note that a constructor generalises the notion of catalyst in resource theory \cite{COC}, relaxing the requirement to stay in exactly the same state by saying that it has to stay within the same set of states).

For the task $T$ to be possible, in general it is not enough that a constructor exists under the given laws: there should also be no limitation to how well a constructor could be approximated, {\sl given the supplementary conditions} to the dynamical laws (e.g. conditions on the available initial states of the dynamics in non-relativistic quantum theory). 
However, under our assumptions (all the states are permitted in unlimited number), a constructor being allowed implies that the corresponding task is possible.\\


{\sl Results.} We will now provide a simple toy model where the dynamical laws are time-reversal symmetric (unitary quantum theory), but the fact that a task $T= \{\rho_x\rightarrow\rho_y\}$ is possible does not necessarily imply the same for its transpose $T^{\sim}$, demonstrating the compatibility between time-reversal symmetric laws and constructor-based irreversibility.

\begin{figure}[ht]
\begin{center}
\includegraphics[width=0.99\columnwidth]{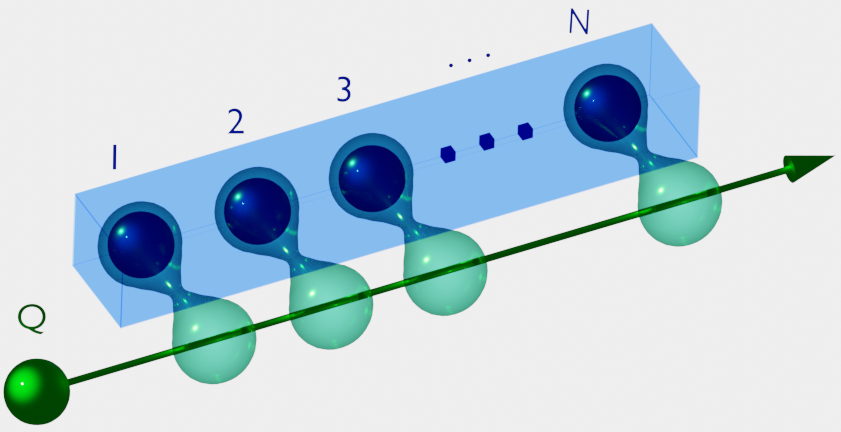}
\caption{Qubit-based homogenisation machine. A substrate qubit $Q$ (in green) interacting with a homogenisation machine (azure shaded box) composed of $N$ qubits (in blue). The emerald hourglasses represent the unitary partial swaps $U_{Q,k}$ ($k=1,...,N$).}
\label{homogeniser}
\end{center}
\end{figure}

This model utilises the powerful scheme of quantum homogenisation \cite{SCA1, SCA2}, depicted in Fig. \ref{homogeniser} and defined as follows.
Consider $N$ qubits, each prepared in the state $\rho_y$, forming the $N$-qubits set $H_{N}[T]$.
Suppose the substrate $Q$ is initialised with the state $\rho_x$, and then it interacts with the qubits in $H_{N}[T]$ one at a time via the unitary transformation:
\begin{equation}\label{UQk}
  U_{Q,k}= \cos(\eta)I+i\sin(\eta)\Sigma_{Qk}\,
\end{equation}
where $\Sigma_{Qk}$ is the SWAP gate acting on the substrate qubit $Q$ and the $k$-th rest qubit in the homogenisation machine, and $I$ is the identity.
The SWAP is defined as the gate $\Sigma_{12}$ with the property that $\Sigma_{12}\ket{\psi}\ket{\phi}=\ket{\phi}\ket{\psi}, \forall \ket{\psi}\ket{\phi}$.
This $U_{Q,k}$ is a {\sl partial swap}, whose intensity is defined by the real parameter $\eta$: the closer $\eta$ is to $\frac{\pi}{2}$, the closer this transformation is to a standard SWAP. For small $\eta$ values, it is a way of slightly modifying the original state of $Q$, making it closer and closer to the desired output state $\rho_y$.

{The state of the qubit $Q$ after interacting with $H_{N}[T]$ is:
\begin{equation}\label{rhoQN}
 \rho_{Q,N}=\mathrm{Tr}_{1...N} \left[\tilde\prod_{k=N}^{1}U_{Q,k}\left(\rho_x\otimes\rho_y^{\otimes N}\right)\tilde\prod_{k=1}^{N}U_{Q,k}^{\dagger}\right]\;,
\end{equation}}
where $\tilde \prod$ denotes an ordered product.
%
Define now the error in performing the task on $Q$ as:
\begin{equation}\label{epsilon}
 \epsilon_N=1-F(\rho_{Q,N},\rho_y)\;.
\end{equation}
%
One can show \cite{SCA1, SCA2} that $\epsilon_N$ tends to zero as the number of qubits $N$ in the machine tends to infinity:
\begin{equation}\label{limNepsilon}
 \lim_{N\rightarrow \infty}\epsilon_N=0\;.
\end{equation}
%

In other words, the machine $H_{N}[T]$ tends to be capable of performing the task $T$ perfectly when $N$ is large, thus satisfying condition (i) for a constructor.
This is true for any task $T$ transforming any quantum state $\rho_x$ into any other state $\rho_y$. However, not all quantum homogenisation machines satisfy condition (ii), failing to be constructors; hence, not all tasks are necessarily possible.
Specifically, $T^{\sim}$ need not be possible, even if $T$ is possible: this is how constructor-based irreversibility emerges.
Note how this trait makes it radically different from the standard irreversibility of the homogeniser, analysed in \cite{SCA1, SCA2} and demonstrated by various qubit implementations (see e.g. \cite{VIBH} for a recent one).

Consider the special case where $\rho_x$ and $\rho_y$ are, respectively, a pure and a maximally mixed state.
%
%
In this case, the task $T$ goes in the direction of more mixedness, while $T^{\sim}$ goes in the opposite direction, purifying the state.
For small $\eta$, it is possible to show two facts (see Supplemental Material for details):

1) As $N$ increases, the homogenisation machine $H_{N}[T]$ tends to be a constructor for $T$, because the relative deterioration after performing the task once goes to 0:
\begin{equation}\label{limT}
  \lim_{N \rightarrow \infty}\lim_{n\rightarrow \infty}\delta_{H_{N}[T]}(n)=0\;.
\end{equation}
%

2) The optimal candidate to perform $T^{\sim}$, $H_{N}[T^{\sim}]$, is {\sl not} a constructor for $T^{\sim}$. Specifically, one can show that
\begin{equation}\label{limTsim}
  \lim_{N \rightarrow \infty}\lim_{n\rightarrow \infty}\delta_{H_{N}[T^\sim]}(n) \rightarrow \infty \;.     
\end{equation}
%
Thus, $T$ being possible and the assumption of time-reversal symmetric laws {\sl do not imply} that $T^{\sim}$ must also be possible. This makes constructor-based irreversibility compatible with time-reversal symmetric laws under unitary quantum theory.\\



\begin{figure}[htbp]
\begin{center}
\includegraphics[width=0.7\columnwidth]{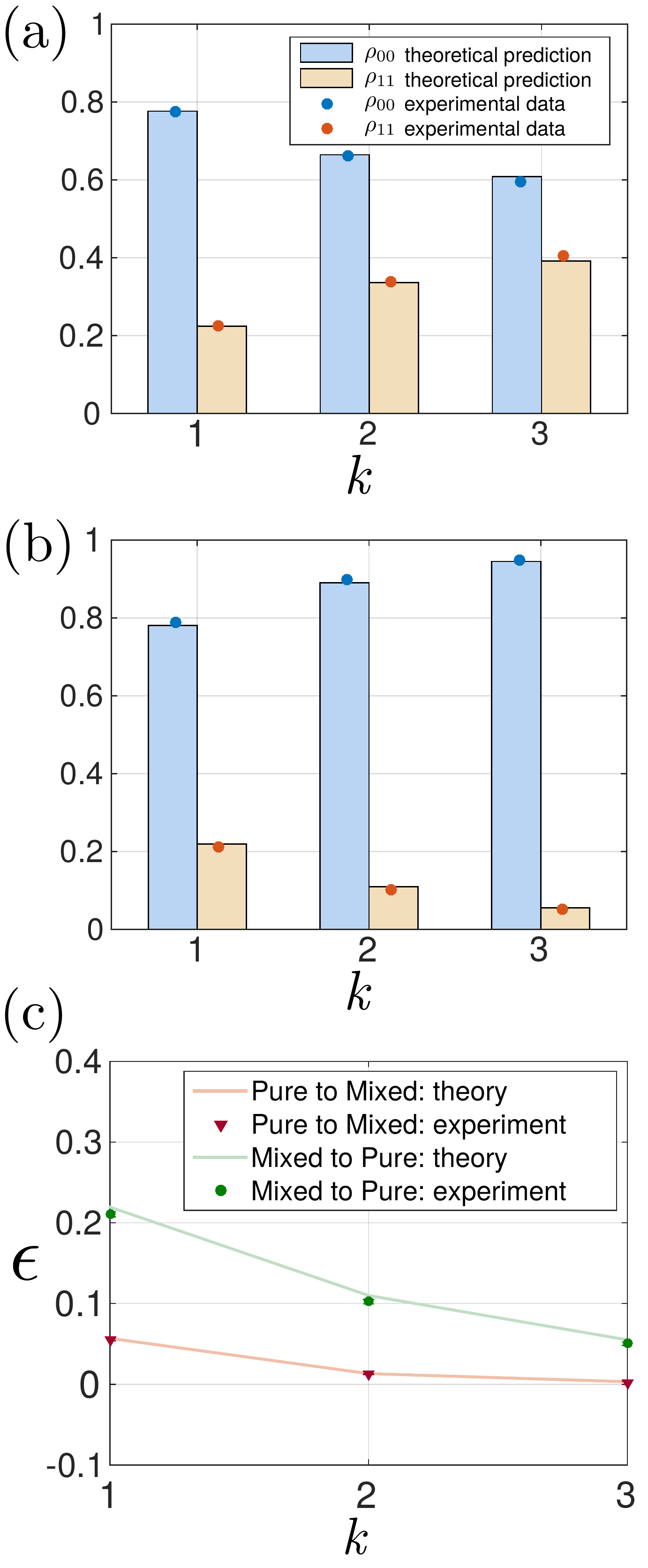}
\caption{Results obtained with partial swap parameter $\eta=\frac{\pi}{4}$, for the first usage of the homogenisation machine ($n=1$). Panel (a): pure-to-mixed task $T$. The plot hosts the $\rho_{00}$ (left side, in azure/blue) and $\rho_{11}$ (right side, in yellow/red) elements of the substrate $Q$ density matrix, initially in the pure state $\rho_p=|0\rangle\langle0|$, after each partial swap $U_{Q,k}$ with one of the $N=3$ homogenisation machine qubits in the mixed state $\rho_m=0.55|0\rangle\langle0|+0.45|1\rangle\langle1|$ (experimentally measured), showing the progression with $k$ of the substrate state evolution induced by the machine. The bars represent the theoretical predictions, while the dots indicate the experimentally-reconstructed values, reported with the associated experimental uncertainties. Panel (b): mixed-to-pure task $T^{\sim}$. The plot structure is the same as the one in panel (a), but here the substrate initial state is $\rho_m=0.55|0\rangle\langle0|+0.45|1\rangle\langle1|$, while the homogenisation machine qubits are in the pure state $\rho_p=|0\rangle\langle0|$. Panel (c): error $\epsilon$ between the task target state and the substrate state undergoing the task (for $k=1,...,N$), both for $T$ (in orange) and $T^\sim$ (in green). Lines: theoretical expectations. Dots: experimental values.}
\label{BS5050}
\end{center}
\end{figure}

{\sl The experiment.} To provide evidence of this mechanism at work, we verified experimentally that the homogenisation machine performing the task $T=\rho_p \rightarrow \rho_m $ (being $\rho_p$ a pure state and $\rho_m$ a mixed one) is more efficient than the machine performing the transpose task $T^{\sim}=\rho_m \rightarrow \rho_p${, even at the first task execution ($n=1$).}
This represents a clear demonstration of the convergence properties mentioned above, showing that, even though the underlying dynamics is time-reversal symmetric, $T^{\sim}$ may not be possible even if $T$ is possible.
Let us consider two states: the pure state $\rho_p=\ket{0}\bra{0}$ and the quasi-maximally mixed state $\rho_m=\frac{1+\gamma}{2}|0\rangle\langle0|+\frac{1-\gamma}{2}|1\rangle\langle1|$ (with $\gamma\ll1$, to take into account experimental imperfections in the mixture preparation).
We compare the performance of the machine $H_{N}[T]$, with the substrate prepared in the state $\rho_p$ (pure to mixed task $T$), with the one of the machine $H_{N}[T^{\sim}]$, acting on a substrate in the state $\rho_m$ (mixed to pure task $T^\sim$), by measuring the accuracy of each machine in performing its task (i.e. the error $\epsilon$).
This is achieved in an experiment (Fig. \ref{setup}) based on qubits realised with single photons at 1550 nm, see Supplemental Material for a detailed description of the setup.
\begin{figure}[htbp]
\begin{center}
\includegraphics[width=0.99\columnwidth]{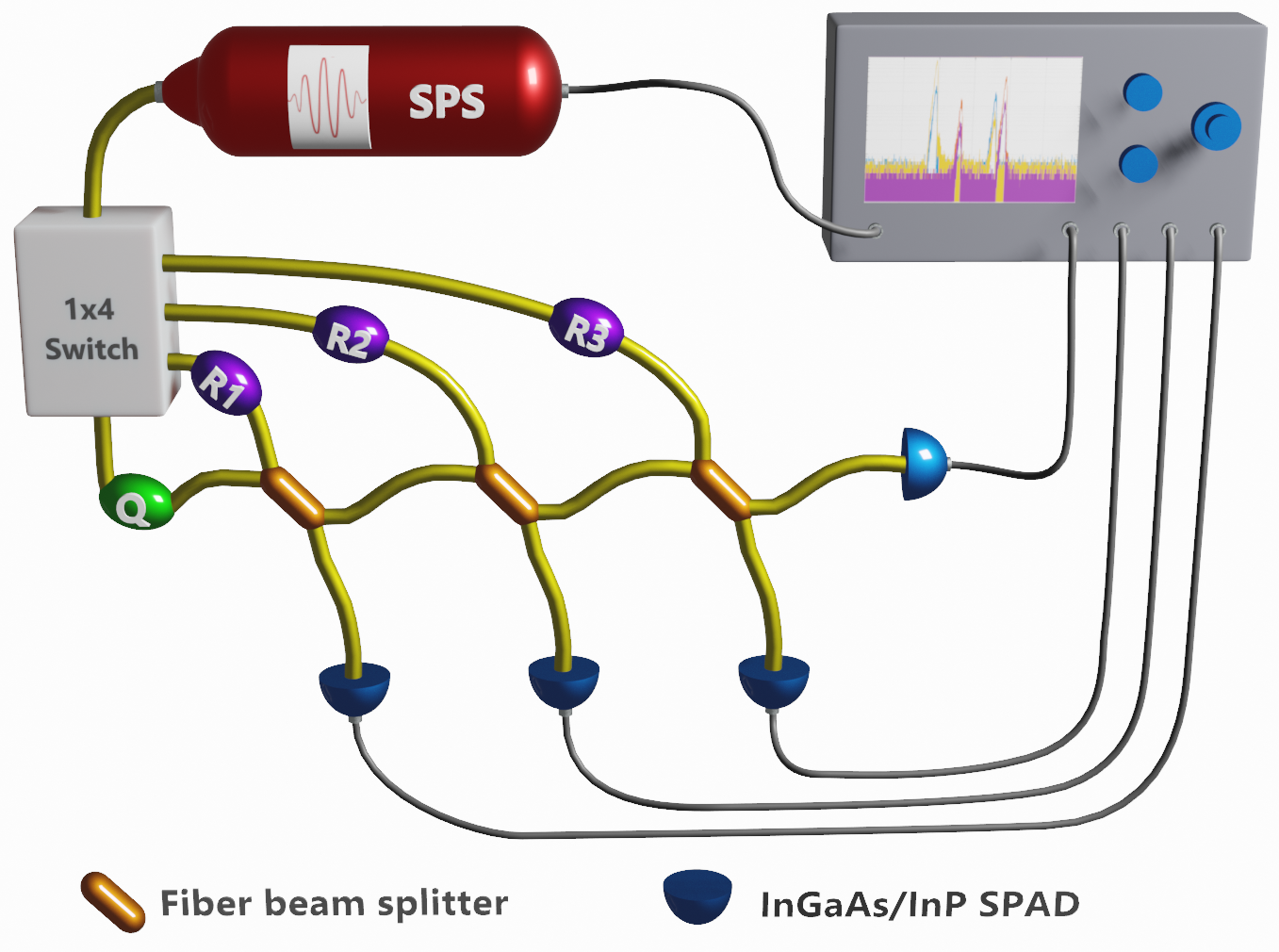}
\caption{{Experimental setup. Heralded single photons at 1550 nm are produced by a low-noise prototype of single-photon source \cite{sorgOE,sorgAPL}. The heralded photon goes to a $1\times4$ fiber optical switch, addressing it either to the substrate path ($Q$) or to one of our 3-qubit homogenisation machine paths ($R1$, $R2$, $R3$). The heralded single photons then meet a cascade of $N=3$ fiber beam splitters, each responsible for one of the partial swaps implemented, simulating the interaction between the substrate and the homogenisation machine. Finally, the photons are detected by free-running infrared detectors, whose output is sent to a time-tagging electronics together with the heralding counts.}}
\label{setup}
\end{center}
\end{figure}
Fig. \ref{BS5050} shows the results obtained for a partial swap parameter $\eta=\frac{\pi}{4}$.

Plots (a) and (b) report, respectively, the progression of tasks $T$ and $T^{\sim}$ as the substrate $Q$ interacts with the $k=1,...,N$ qubits of the corresponding homogenisation machines, in the case $n=1$.
The reconstructed diagonal elements $\rho_{00}$ and $\rho_{11}$ of the $Q$ density matrix are reported (by construction, in our case $\rho_{01}=\rho_{10}=0$), considering for $T$ the substrate in the initial state $\rho_p=|0\rangle\langle0|$ and the homogenisation machine qubits in the mixed state $\rho_m=0.55|0\rangle\langle0|+0.45|1\rangle\langle1|$, and viceversa for $T^{\sim}$.
The experimentally-reconstructed values of $\rho_{00}$ and $\rho_{11}$ (blue and red dots, respectively) are in very good agreement with the theoretical predictions (azure and yellow bars, respectively) throughout the whole process in both cases.
Plot (c) shows the error $\epsilon$ in Eq. (\ref{epsilon}), indicating the discrepancy between the $Q$ density matrix and the target state one.
The experimental points, in good agreement with the theoretical expectations (solid lines) for both tasks, show how the homogenisation machine for $T$ always outperforms the one for $T^{\sim}$.
%
\begin{figure}[h!tbp]
\begin{center}
\includegraphics[width=0.99\columnwidth]{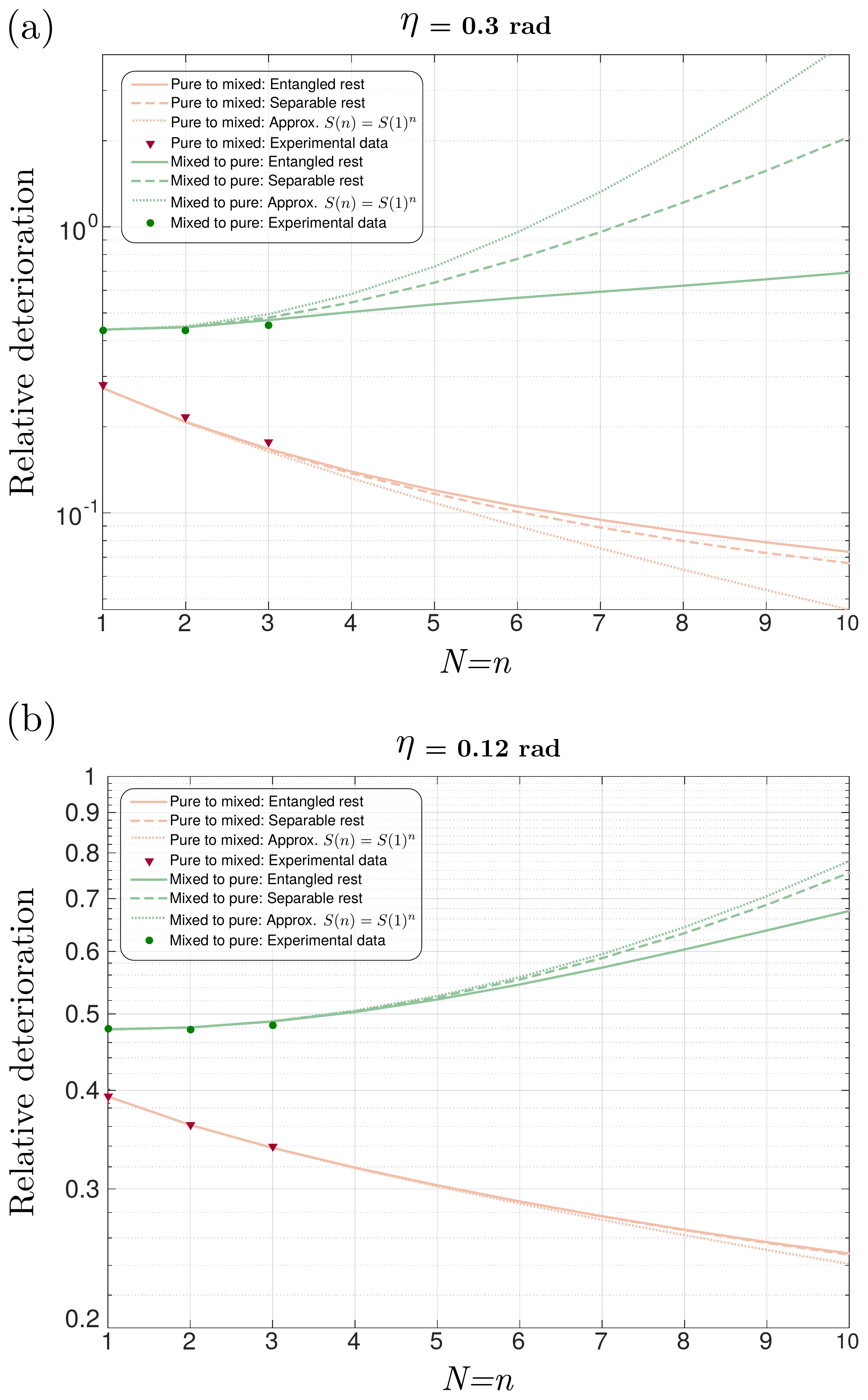}
\caption{Constructor-based irreversibility demonstration. Plots (a) and (b) show, for $\eta=0.3$ and $\eta=0.12$ respectively, the relative deterioration of the homogenisation machine for task $T$ and its counterpart $T^{\sim}$, as a function of the number of machine qubits $N$ and machine usages $n$. The solid curves represent the theoretically-expected behavior considering the entangled rest state of the complete theory for $n>1$, while the dashed curves show the predictions given by approximating the homogeniser for $n>1$ as an ensemble of qubits in a separable state. The dotted curves show, instead, the approximation considering for each $j=1,...,n$ a separable rest $R$, in the limit of small $\eta$ values and with $S_{{\cal E}[T]}(n)\approx S_{{\cal E}[T]}(1)^{n}$. Finally, the triangles (dots) indicate the experimental values obtained for task $T$ ($T^{\sim}$).}
\label{deterioration}
\end{center}
\end{figure}
%
In addition, we extended our analysis to the general case $n>1$, to verify the behavior of the relative deteriorations $\delta_{H_{N}[T]}(n)$ and $\delta_{H_{N}[T^\sim]}(n)$ (see Eq. (\ref{reldet}), and Supplemental Material for details).
To this end, we evaluated the machines steadiness with a recursive method considering, for the $j$-th substrate state $Q_j$ ($j=1,...,n$), the machine initialized in a state as close as possible to the one of the machine outgoing the $(j-1)$-th substrate-machine interaction.
For each $j$, after executing the task and tracing over the substrate $Q_j$ in Eq. (\ref{rhoRn}), the state $\rho_R^{(j)}$ includes some entanglement between the qubits of the rest.
Although we were not able to directly measure and quantify this entanglement, we could nevertheless observe the resulting correlations among the rest qubits, so we reconstructed these correlations (in the computational basis $\{|0\rangle,|1\rangle\}$) and reproduced them while initializing the machine state for the $(j+1)$-th usage.
The results for both tasks $T$ and $T^{\sim}$ are reported in Fig. \ref{deterioration}.

{The figure illustrates, for the two experimental $\eta$ values corresponding to a weak interaction between substrate and homogenisation machine, the behavior of the relative deteriorations $\delta_{H_{N}[T]}(n)$ and $\delta_{H_{N}[T^\sim]}(n)$, providing an evidence of their asymptotic limits reported in Eqs. (\ref{limT}) and (\ref{limTsim}).\\
By performing numerical simulations, we studied the relative deterioration behavior for a limited number $N$ of rest qubits and $n$ machine usages, extending our analysis beyond the $N=3$ experimental limit.
For the sake of completeness, we investigated both the ideal scenario, with the homogeniser qubits in an entangled state for $n>1$ (solid lines), and the case of the rest qubits forming a separable state for each use of the homogenisation machine (dashed lines).
We also showed the relative deterioration behavior (dotted lines) in the approximation of small swap intensity $\eta$ and a homogenisation machine steadiness at the $n$-th cycle given by $S_{{\cal E}[T]}(n)\approx S_{{\cal E}[T]}(1)^{n}$, with $\rho_R^{(n)}$ taken as the tensor product of the reduced density operators of the rest qubits emerging from the $n$-th task execution.
Such approximation allows finding an analytical solution for $\delta_{H_{N}[T]}(n)$ and $\delta_{H_{N}[T^\sim]}(n)$, satisfying the conditions in Eqs. (\ref{limT}) and (\ref{limTsim}), respectively (details in Supplemental Material).\\
Our experimental results, obtained evaluating the fidelities of the detected events in the computational basis, are in good agreement with the theoretical predictions for the entangled rest scenario.
Fig. \ref{deterioration} shows that, for the pure-to-mixed task $T$, the relative deterioration $\delta_{H_{N}[T]}(n)$ steadily decreases for increasing values of $n,N$, qualifying the homogenisation machine $H_{N}[T]$ as a proper constructor for the task $T$, according to Eq. (\ref{limT}).
Conversely, we observe that $\delta_{H_{N}[T^\sim]}(n)$ diverges for increasing $n$ and $N$, in agreement with the scenario corresponding to Eq. (\ref{limTsim}), allowing us to state that the homogeniser $H_{N}[T^\sim]$ fails to be a constructor for $T^\sim$.
Hence we can conclude that, in the constructor theory framework, the task $T$ is ``possible'', while its counterpart $T^\sim$ need not be: this is what makes the process corresponding to the task $T$ potentially irreversible.\\

{\sl Conclusions.} In this paper we proposed a novel take on the old-age problem of reconciling irreversibility with reversible unitary dynamics with a radically different approach, considering the irreversibility based on possibility/impossibility of tasks rather than on dynamical trajectories being permitted or disallowed.
This notion of irreversibility extends the one of classical thermodynamics to a general information-theoretic scenario, thus representing a significant contribution to the development of a quantum thermodynamics \cite{ade21,horo13,and16,THD1,kur08}.
The problem is addressed by using the conceptual framework of constructor theory, where the fact that a certain task $T$ is possible does not imply that the transpose task $T^\sim$ is possible too.
Here we have demonstrated this idea with a specific example (exploiting homogenisation machines), also providing an experimental demonstration of this mechanism at work.
As shown by the results obtained in our single-photon experiment, the homogenisation machine implementing the pure-to-mixed task $T$ always outperforms its counterpart for the reverse task $T^{\sim}$. Furthermore, by looking at the relative deterioration of both machines it is evident how the one for $T^{\sim}$ suffers much higher degradation than the one realising $T$, ultimately not satisfying condition (ii) and thus failing to qualify as a proper constructor: this gives a clear proof of the compatibility of the constructor-based irreversibility with unitary quantum theory, providing also a frame for the emergence of a thermodynamical irreversibility in quantum mechanics \cite{ghosh,uzd,lost,riech,scar,manz}.
These results have also significant application in studying quantum computational processes and their reversibility.

\section{Acknowledgments}
We are grateful to David Deutsch, Benjamin Yadin, Paul Raymond-Robichaud and Maria Violaris for useful discussions.
VV thanks the Oxford Martin School, the John Templeton Foundation, the EPSRC (UK).
LK and FP thank Margot Gramegna for helping with the experimental setup optimisation.\\
This work has received funding from the European Union's Horizon 2020 and the EMPIR Participating States in the context of the project EMPIR-17FUN01 ``BeCOMe'', and by the European Union's Horizon 2020 FET-OPEN project grant no. 828946 ``Pathos''.
This research was supported by the grant number (FQXi FFF Grant number FQXi-RFP-1812) from the Foundational Questions Institute and Fetzer Franklin Fund, a donor advised fund of Silicon Valley Community Foundation.
This research was also supported by the National Research Foundation, Prime Minister’s Office, Singapore, under its Competitive Research Programme (CRP Award No. NRF- CRP14-2014-02) and administered by Centre for Quantum Technologies, National University of Singapore.
CM's research was also supported by the Templeton World Charity Foundation and by the Eutopia Foundation.

\end{document}